\documentclass[10pt,a4paper,twocolumn]{article}

\usepackage{amsmath,amssymb}
\usepackage{multicol}

\usepackage{changepage}
\usepackage{units}
\usepackage{graphicx}
\usepackage[margin=1in]{geometry}
\usepackage{sidecap}
\sidecaptionvpos{figure}{t}

\usepackage[utf8x]{inputenc}
\usepackage{cite}

\usepackage{nameref,hyperref}

\usepackage{xcolor}

\usepackage{array}

\usepackage{siunitx}
\usepackage{booktabs}

\makeatletter
\renewcommand{\@biblabel}[1]{\quad#1.}
\makeatother

\date{}

\usepackage{epstopdf}
\begin{document}
\twocolumn[
\begin{@twocolumnfalse}

\begin{flushleft}
{\Large
\textbf\newline{Correlated microtiming deviations in jazz and rock music} 
}
\newline
\\
Mathias Sogorski\textsuperscript{1,3},
Theo Geisel\textsuperscript{1,2,3},
Viola Priesemann\textsuperscript{1,2}
\\
\bigskip
\textbf{1} Max Planck Institute for Dynamics and Self-Organization, Göttingen, Germany \\
\textbf{2} Bernstein Center for Computational Neuroscience, Göttingen, Germany\\
\textbf{3} Department of Physics, Georg-August University, Göttingen, Germany\\

\end{flushleft}

\section*{Abstract}

Musical rhythms performed by humans typically show temporal fluctuations. While they have been characterized in simple rhythmic tasks, it is an open question what is the nature of temporal fluctuations, when several musicians perform music jointly in all its natural complexity. To study such fluctuations in over 100 original jazz and rock/pop recordings played with and without metronome we developed a semi-automated workflow allowing the extraction of cymbal beat onsets with millisecond precision.
Analyzing the inter-beat interval (IBI) time series revealed evidence for two long-range correlated processes characterized by power laws in the IBI power spectral densities. One process dominates on short timescales ($t < 8$ beats) and reflects microtiming variability in the generation of single beats. The other dominates on longer timescales and reflects slow tempo variations.
Whereas the latter did not show differences between musical genres (jazz vs. rock/pop), the process on short timescales showed higher variability for jazz recordings, indicating that jazz makes stronger use of microtiming fluctuations within a measure than rock/pop.
Our results elucidate principles of rhythmic performance and can inspire algorithms for artificial music generation.
By studying microtiming fluctuations in original music recordings, we bridge the gap between minimalistic tapping paradigms and expressive rhythmic performances.

\vspace{1.3cm}

\end{@twocolumnfalse}
]

\section*{Introduction}


The art of creating music involves a balance of surprise and predictability. This balance needs to be achieved on many scales, and for many musical components like melody, dynamics, and rhythm.
Such a balance is believed to be essential for making music interesting and appealing \cite{Huron2006,Levitin2012,Hennig2012,Schmidhuber2008, Rasanen2015}. 
While musicians achieve this balance intuitively, the principles generating it remain unknown. 
A core hypothesis conjectures that this balance manifests itself in long-range correlations (LRCs) and self-similar structure of melody, dynamics, and rhythm.
In fact, a first evidence for this hypothesis was provided by Voss and Clarke \cite{Voss1975}, who identified LRCs in pitch and loudness fluctuations.
More recently, LRCs were found in the rhythmic structure of Western classical music compositions\cite{Levitin2012}, i.e. in
written notations, where the rhythm is represented in a metrically organized precise fashion.
Such compositions may be played back in this precise fashion, e.g., by computers, but are often perceived to sound mechanical and unnatural \cite{Hennig2011}. 
In performed music, in contrast, musicians introduce subtle deviations from the metrically precise temporal location, which make the performance sound human.
\begin{figure*}[h!tb]
	\centering
	\includegraphics[width=.8\textwidth]{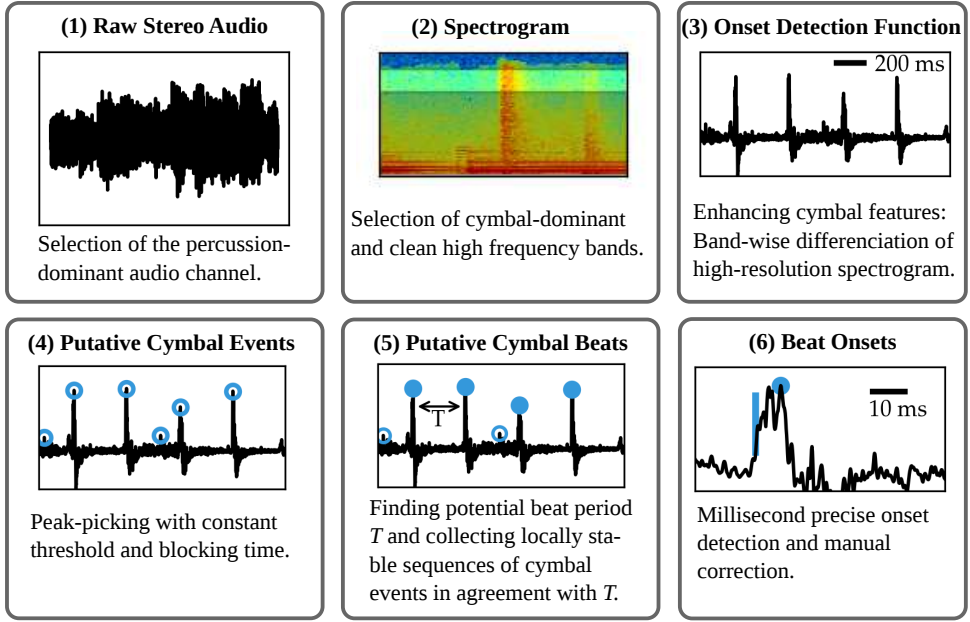}
	\caption{Workflow for the estimation of the beat-onset time series from one channel of a music recording.}
	\label{fig:Pipeline}
\end{figure*}
Such \emph{microtiming deviations} on the one hand are inevitable in human performances as human abilities to produce precisely timed temporal intervals are limited \cite{Repp2013,Merchant2008}.
On the other hand, they can be introduced on purpose and contribute to a musician's individual expression. It is thus worthwhile elucidating the nature of temporal fluctuations and factors contributing to them in various musical contexts.
Inferring such microtiming deviations from ready-made musical recordings is a challenge, however, because beat onsets must be determined with millisecond precision.
In past studies this precision was achieved using fairly reduced settings, e.g. simple finger-tapping tasks
\cite{Hennig2011,Ding2002,Torre2008,Smit2013, 
 Wing1973b,Delignieres2009,
Torre2009, Gilden1995,Chen2002,Delignieres2004,Musha1985,Hennig2014}. 
For those performed with metronome, LRCs were identified for \emph{microtiming deviations} ($e_i$) from  metronome clicks \cite{Hennig2011,Ding2002,Torre2008,Smit2013}.
Here, LRCs manifest themselves as power-laws $P(f)\propto f^{-\beta}$, with $0.5 \lesssim \beta \lesssim 1.5$ in the power spectral density (PSD) of the $e_i$.
In contrast, if the deviations $e_i$ were independent, one would expect $\beta=0$.
For unpaced tapping, i.e. tasks performed without a metronome, LRCs were recovered for \emph{tempo fluctuations}, i.e. the PSD of the \emph{inter-beat interval} (IBI) time series showed  power-laws $P(f)$ with  $0.5 \lesssim \beta \lesssim 1.5$ \cite{Torre2008,Torre2009, Hennig2011,Gilden1995,Chen2002,Delignieres2004,Musha1985,Hennig2014}. 
Hennig and colleagues extended this framework to more complex rhythms (with metronome), but still in a laboratory setting, They provided evidence for LRCs of microtiming deviations, consistently with those of simple finger tapping \cite{Hennig2011}.
More recently, LRCs were identified for drumming in a single pop song \cite{Rasanen2015}.
Together, these results may suggest that both, microtiming deviations from beats, as well as tempo fluctuations show LRCs.
Detailed analyses are required to investigate this hypothesis, in particular with respect to the precise scaling properties (i.e. $\beta$), and their dependence on genres.
Differences in scaling may occur, as the cognitive involvement clearly differs between simple tapping tasks versus the flow experienced when making music together.

In our present study, we carry the analysis of human beat performance from the laboratory to real-world conditions of musical performances with all its complexity. 
To this end, we compiled beat onset time series from over 100 music recordings.
To estimate the beat onset for each recording with millisecond precision, we devised a semi-automated beat extraction workflow.
The resulting IBI time series allowed us to investigate both, unpaced and paced recordings, and to compare their scaling properties to those from finger tapping.
Making use of our large dataset, we extended our analysis to investigate different genres, jazz and rock/pop,  to elucidate how genre-dependence manifests itself in the beat structure.

Based on the millisecond precise beat time series, we could identify signatures of two processes, a clock and a motor process.
Both processes influence the beat microtiming and showed similar long-range correlations. However, the motor process revealed stronger timing fluctuations within a measure for jazz compared to rock/pop. On the one hand our results point to general dynamics of microtiming fluctuations across musical genres on long time scales, reflecting the temporal organization of  musical pieces. On the other hand the stronger fluctuations on fast time scales in jazz music might be attributed to the higher degree of freedom as compared to rock/pop.

\section*{Results}

\subsection*{Millisecond-precise beat extraction}

Human rhythmic performance can be precise down to the scale of several milliseconds \cite{Repp2013,Merchant2008}.
Therefore, our analyses required a millisecond-precise, consistent estimation of beat onsets.
As this precision is not reached by any of the currently available methods,
we developed a specialized semi-automated workflow.

A  conceptual challenge in beat detection of original performances is that the beat is not uniquely defined.
We approximated the beat by cymbal onsets, because drummers provide a rhythmic foundation, because cymbal onsets can be well separated from other instrument onsets, and because the short attack times allow for millisecond-precise onset detection. 
This precise onset detection is crucial for the subsequent systematic and reproducible analyses of large datasets.

In the following, we sketch the semi-automated workflow for beat-extraction (see Fig~\ref{fig:Pipeline}). 
More details are given in the methods section.
(1) The percussion-dominated channel is selected. 
(2) The frequency range in which the cymbal dominated is isolated.
(3,4) Using differentiation, putative  cymbal events are identified. 
(5) Of those, the cymbal onsets that built a regular beat sequence are combined to a beat-onset time series. This step excludes cymbal onsets that were not on the regular beat. 
(6) To improve the temporal precision of the extracted beat onsets, the precise onset time is estimated on the rising slope of the cymbal beat.
This workflow allowed us to acquire beat time series from more than 100 recordings, comprising each about 600 beat onsets. All songs we analyzed are listed in 

\subsection*{Human beat performance in music}

\begin{figure}[h!tb]
	\includegraphics[width=\columnwidth]{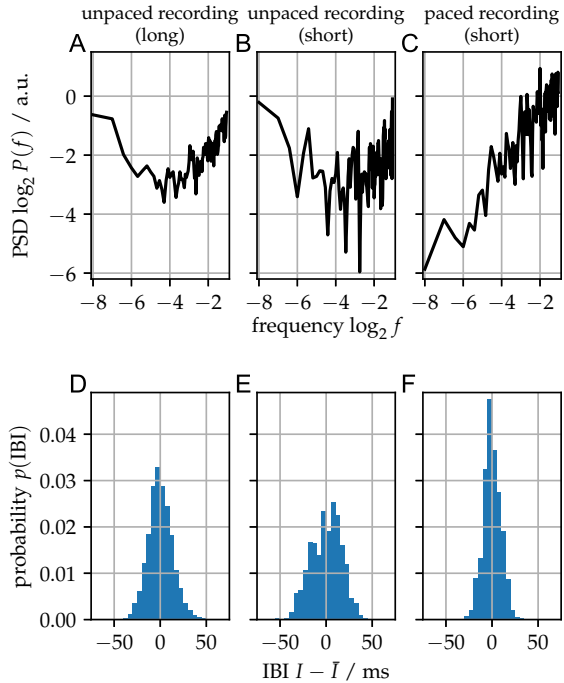}
	\caption{A-C. Power spectral densities (PSD) of inter-beat intervals (IBIs) for unpaced and paced (metronome-guided) recordings. ``Long'' refers studio drum recordings  of about 30~min and ``short'' to jazz and rock/pop recordings of typically $\approx 3$~min (A: long jazz recording; B: Buster Smith -- Kansas City Riffs; C: Bee Gees - How Deep Is Your Love).
	D-F. IBI distribution for the same example recordings.}
	\label{fig:Spectral-Shapes}
\end{figure}

\begin{figure*}[h!]
	
		\centering
		\includegraphics[width=\textwidth]{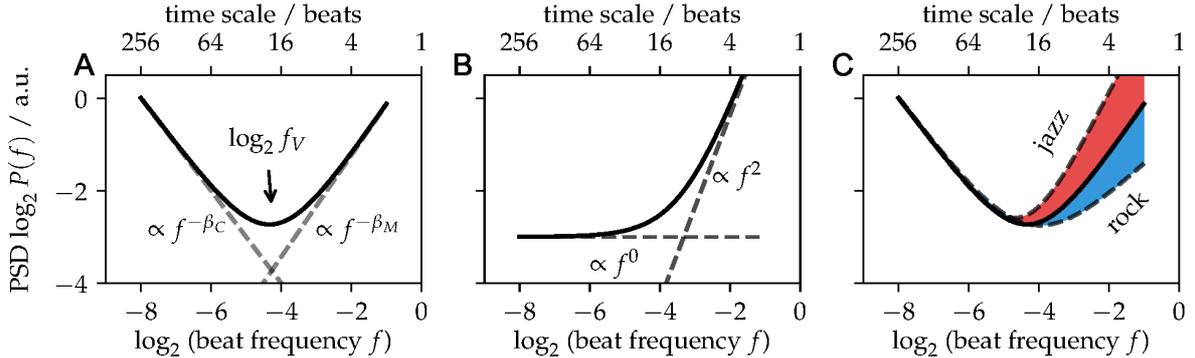}
		
		\caption{Scheme of the power-spectral density (PSD) of inter-beat interval (IBI) time series. 
			A. The combination of a long-range correlated clock and motor process (dashed lines) superpose to a V-shaped PSD of the IBIs (solid line). The characteristic frequency (or time scale, see upper axis) at the minimum determines the turnover between the clock-dominated and motor-dominated regime.
			B. Same as panel A, but assuming that both, the clock and motor process are uncorrelated.
			C. Sketch of genre-induced differences in the PSD.}
		\label{fig:V-shape-Scheme}

\end{figure*}

\begin{figure*}[h!t]
	\centering
	\includegraphics{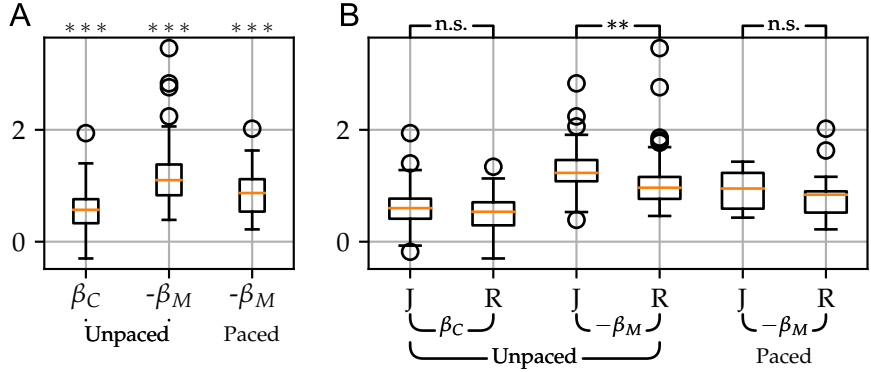}
	\caption{Scaling exponents ($\beta$) obtained from analyzing the PSDs of IBI time series. 
	A.~Neither the clock nor the motor process is random but clearly long-range persistent, i.e. $\beta_C > 0, -\beta_M < 2$. *** denotes $p\ll 10^{-3}$ (significance obtained by bootstrapping).
	B.~Genre-dependence of the scaling exponents. The motor process showed significantly  stronger long-range persistence in rock/pop (R) than in jazz (J). The box plot depicts the median in red, boxes at the first and third quartile, whiskers at $1.5\cdot\mathrm{IQR}$ (interquartile range), and circles represent outliers. 
	** denotes $p=0.001$, and n.s. denotes \textit{not significant}.
	}
	\label{fig:Boxplots}
\end{figure*}

We analyzed recordings played with or without metronome.
For those played with a metronome (paced recordings), we  consistently found a power law for the power spectral density (PSD) of the inter-beat intervals (IBIs) (Fig~\ref{fig:Spectral-Shapes}C, sketch in Fig~\ref{fig:V-shape-Scheme}). 
The exponents $\beta_M$ of the motor or microtiming deviations varied across recordings, but consistently indicated long-range correlations (LRCs).
Its median was $\bar{\beta}_M = \num{-0.87\pm 0.43}$ (where the standard deviation is given in parentheses).
$\bar{\beta}_M$ is negative, because the IBI time series, compared to the deviations from the metronome, represents a differentiated signal (see below). $\bar{\beta}_M$ significantly differed from an independence assumption ($\beta_M^\mathrm{ind}=-2$, $p<10^{-30}$,  where significance was obtained by analytical calculation of the bootstrap distribution; Fig~\ref{fig:Boxplots}A).
Qualitatively, these results are consistent with those for simple finger-tapping tasks, indicating that a similar process underlies beat generation in simple tapping tasks as well as in music.

IBI time series from  \textit{unpaced} music recordings showed characteristic V-shapes for the PSD (Fig~ \ref{fig:Spectral-Shapes}A,B).
Such V-Shapes can be generated by the superposition of two stochastic processes, each of them contributing to the PSD.
In analogy to finger-tapping experiments, we interpret the two processes as a ``clock process'' $C$ governing temporal interval estimation, and a ``motor process'' $M$ governing the motor execution of a planned interval (Fig~\ref{fig:V-shape-Scheme}A) \cite{Wing1973b,Torre2009,Gilden1995}.
In this general framework, an IBI interval $I_i$ is generated by a clock estimate $C_i$, and motor deviations $M_i$, which represent the microtiming deviations from the intended clock interval $C_i$ \cite{Wing1973b, Gilden1995, Delignieres2004, Torre2009}:
\begin{align}
	I_i = C_i + M_i − M_{i−1}, \label{eq:Gilden-TS}
\end{align}
The PSD of the intervals $I$ is thus generated by the PSDs of the two stochastic processes, $C$ and $M$.
The clock process contributes with a power law  $P(f)\propto f^{-\beta_C}$, where $0.5 \lesssim \beta_C \lesssim 1.5$ for long-range correlations.
For an uncorrelated process one expects $\beta_C^\mathrm{ind}=0$ (Fig~\ref{fig:V-shape-Scheme}A,B).
The motor process $M$ enters $I$ as a difference, and hence contributes to the PSD with $-1.5 \leq \beta_M \leq -0.5$ for long-range correlations, whereas $\beta_M^\mathrm{ind}=-2$ would reflect an uncorrelated process (Fig~\ref{fig:V-shape-Scheme}A,B).
As the clock and motor processes contribute to $I$ with exponents of opposite sign, $\beta_C>0$ and $\beta_M<0$, respectively, $C$ dominates the PSD at low frequencies,  whereas $M$ dominates at high frequencies.
This generates the characteristic V-shape and allows to estimate both scaling exponents from the PSD of the IBIs (Fig~\ref{fig:Spectral-Shapes}A,B, \ref{fig:V-shape-Scheme}A).
When the rhythm is performed with a metronome (paced), the dynamics of the clock process is strongly confined, and the motor process alone dominates the PSD on the entire frequency range, i.e. one observes 
a single power law regime in the power spectral density (PSD) of the IBI time series, as reported above (Fig~ \ref{fig:Spectral-Shapes}C) \cite{Torre2009, Hennig2011}.

\begin{figure*}[h!tb]
	\centering
	\includegraphics[width=\textwidth]{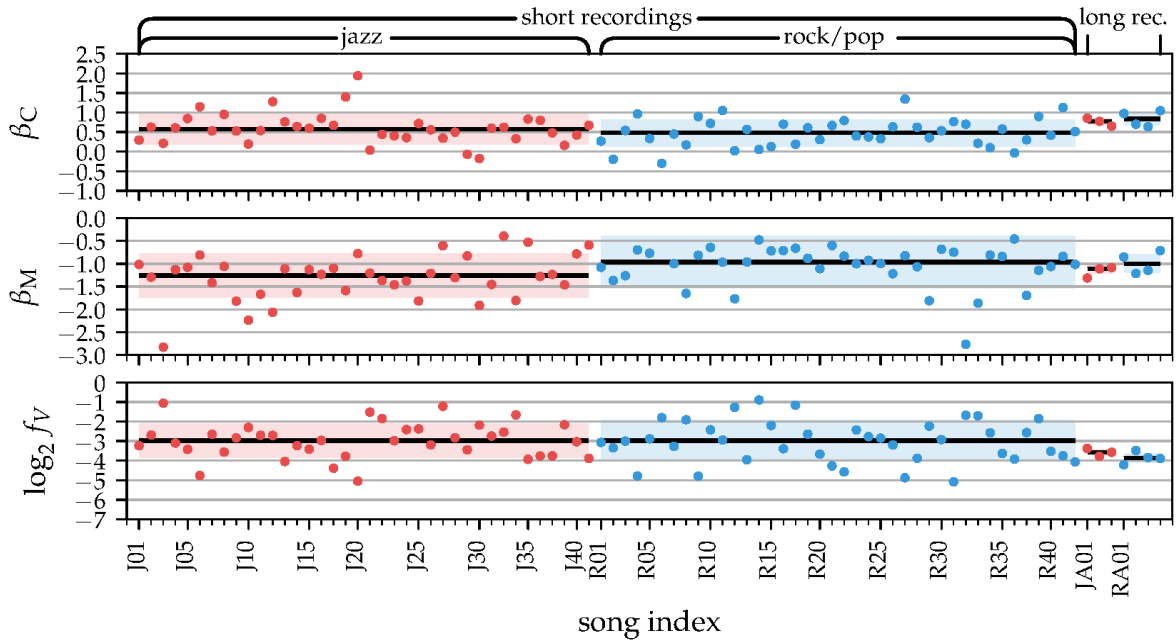}
	\caption{Fit parameters of the unpaced musical recordings by genre (jazz: red, rock/pop: blue). 
		Both, the music performances and our studio recordings showed consistent tendencies (median (SD) given as black lines (colored ranges) for the individual parameters). Most prominently, the motor persistence $\beta_M$ is genre-dependent.}
	\label{fig:Overview}
\end{figure*}

We systematically quantified the scaling exponents $\beta _C$ and $\beta_M$ for \textit{unpaced} recordings  (Fig~\ref{fig:Boxplots},\ref{fig:Overview}).
The clock process showed LRCs with  ${\bar{\beta}_C = \num {0.54\pm 0.38}}$. It significantly differed from an independent process, which would be characterized by $\beta^\mathrm{ind}_C=0$ ($p=10^{-29}$, bootstrap).
Our results indicate that tempo fluctuations across the entire recording do not occur independently, but ultimately are related to fluctuations at any other time.
The motor process contributed with ${\beta_M \approx -1}$ (${\bar{\beta}_M= \num{-1.09\pm 0.55}}$), and significantly differed from an independent process as well (${\beta^\mathrm{ind}_M = -2}$, ${p=10^{-29}}$, bootstrap).

These results can be interpreted as follows: As the local tempo, governed by the clock process, needs to be maintained, any deviation from the local clock or metronome shortens one interval and at the same time lengthens the other, resulting in anti-correlations on $I$, and negative values of $\beta_M$. 
Last, the turnover between the clock- and motor-dominated regimes was generally at about $\log_2 f_V \approx -3$ (${\log_2 \bar{f}_V = \num{-2.98\pm 0.98}}$).
That is, only for about $2^3 = 8$ beats or a few measures the motor process dominated the PSD, while for time scales spanning more than about 8 beats the clock process dominated.

Interestingly, unpaced and paced recordings  only differed slightly in their IBI distributions $p(\mathrm{IBI})$ (Fig~\ref{fig:Spectral-Shapes}D-F). Despite the absence of a metronome, unpaced recordings showed only slightly broader $p(\mathrm{IBI})$ ($\sigma = 13.1(3.8)$~ms and $\sigma = 11.5(4.3)$~ms, respectively, $p=0.067$, $d=0.393$). 
Moreover, for both conditions, the microtiming deviations showed similar scaling properties, i.e. the $\beta_M$ did not differ between the conditions ($p>0.05$).
In contrast, the  characteristic V-shape was clearly present for the PSD of $I$ when recordings were played \textit{without} a metronome, whereas those played \textit{with} metronome showed a single power law, because the metronome presumably suppressed or replaced the clock process (Fig~\ref{fig:Spectral-Shapes}). 
This result supports the hypothesis of two independent processes, one being suppressed when beats are performed under pacing by a metronome.

As many recordings are fairly short (about 3 minutes, median of $580$ beats), the effect of spectral averaging was small and thus the PSDs were noisy.
To obtain PSDs from longer time series, we recorded seven unpaced, genuine drum performances from a professional musician in a studio setup, lasting 20 to 30 minutes each and comprising $3189(612)$ beats.
For these long time series, the PSDs were very clear due to better spectral averaging (Fig~\ref{fig:Spectral-Shapes}A).
The parameters obtained from these PSDs were consistent with those of the short musical recordings analyzed above:
$\bar{\beta}_C = \num {0.77\pm 0.15}$, $\bar{\beta}_M= \num{-1.11\pm 0.19}$, and $\log_2 \bar{f}_V = \num{-3.77\pm 0.26}$.
As the short and long recordings did not differ significantly in any of the parameters, we merged both for the analysis of genre-dependence in the following sections.
Note, that we obtained the same results when we considered the short recordings alone, and the same as a trend for the seven long recordings, which alone, however, would not be numerous enough to reach significance.

\subsection*{Genre-dependence}
Do the scaling properties of the clock and motor process depend on the musical genre or are they a general feature of music?
With the highly precise IBI time series from jazz and rock/pop music we were able to test for genre-dependence.

Most interestingly, we found that jazz recordings showed smaller $\beta_M$  for the unpaced songs than rock/pop recordings (Fig~\ref{fig:Boxplots}B, $p=0.001$, $d=0.509$, restricted permutation test (RPT), for details on the statistical tests see methods).
More precisely, for jazz recordings we found $\bar{\beta}_M=\num{-1.23\pm 0.47}$, and for rock/pop $\bar{\beta}_M=\num{-0.96\pm 0.56}$.
The same trend was observed for the paced songs.
In contrast, jazz and rock did not differ in the clock exponent $\beta_C$ (Fig~\ref{fig:Boxplots}B).
These results indicate that in jazz, musicians make more use of microtiming deviations on very short time scales, i.e. they introduce stronger deviations from the local tempo. 
In rock/pop, musicians play with a more regular beat on these short time scales.
On longer time scales, where the clock process dominates, the tempo variations do not differ between jazz and rock/pop, indicating that the overall musical structure from short motives to long blocks does not differ between these genres, and we hypothesize that other genres might show similar LRCs for the clock process as well.

\subsubsection*{Basic beat variability}
In addition to the very prominent genre-dependence in the motor process, we found that on average the beat in jazz was slightly slower than in rock.
In detail, the median IBI in jazz (rock/pop) was $330$~ms ($288$~ms) for the unpaced recordings, and $400$~ms ($282$~ms) for the paced recordings ($p=0.013$, $d=0.574$, RPT).
More interestingly, the variability (i.e. the standard deviation of $I$) was higher for jazz than for rock; it was $14.3$~ms ($11.9$~ms) for the unpaced performances, and $14.6$~ms ($9.2$~ms) for the paced performances ($p = 0.001$, $d=0.745$, RPT).
It is to be expected, that faster performances show less variability (smaller SD).
To account for this, we compared the tempo-normalized variability of the IBI time series, i.e.  the Fano factor $F = \sigma^2 / \bar{I}$, where $\sigma$ denotes the standard deviation and $\bar{I}$ the median $I$.
Consistently with the differences above, the Fano factor was higher in jazz than in rock across the paced and unpaced recordings ($p=0.016$, $d=0.294$, RPT, Bonferroni corrected for multiple comparisons), although the effect size was smaller.
When analyzing the paced and unpaced recordings separately, both showed the same trend  ($p=0.049$, $d=1.211$ for paced, $p=0.132$, $d=0.182$ for unpaced, Bonferroni corrected for multiple comparisons), however, the effect size was more pronounced for the paced songs.
Together, these result suggests that jazz makes more use of temporal variability, especially when recordings are played with a metronome (paced), i.e. when only the motor, but not the clock process can be used as an expressive component of music.

\section*{Discussion}

Interestingly, our results for rhythm generation in music revealed evidence for the same two underlying processes as inferred from simple finger tapping tasks: a clock and a motor process, both of them long-range correlated.
In both settings, music and finger tapping, the clock process disappeared when a metronome was used, and the remaining motor process showed a single power law with slope around unity.
However, although the LRCs in both settings have similar characteristics, their origin may be very different:
When a piece of music is performed, it has structure on all scales, from motifs, phrases and themes to verses and movements. This structure is reflected in the tempo and is likely to underlie the observed long-range correlations.
Such structure is absent in finger tapping tasks.
Those tasks are somewhat dull for the subjects, and hence it may be that their mind is wandering during tapping. As a consequence, concentration may wax and wane, certainly on many different scales as well, thereby generating LRCs.
Studies relating neural activity to motor precision and perception hint in that direction \cite{Palva2013, Smit2013}. 
Smit et al, for example, showed clear correlation between $\beta_C$ and the scaling exponent derived from neural alpha oscillations \cite{Smit2013}. 
Hence, although the signatures of the beat time series are similar for music and finger tapping, their origin may differ vastly: One lying in the multi-scale structure of musical compositions, which aims at keeping us captivated, the other making our mind wander owing to the dullness of tapping a simple beat for minutes in a row.
Such waxing and waning of concentration and performance can be tested in future studies by simultaneously measuring markers of attention in brain activity or pupil diameter and relating this to the microtiming deviations in finger-tapping tasks; in musical pieces, the microtiming deviations might be related to the structure of each piece.

Interestingly, our results for rhythm generation in music hinted at the same underlying processes as inferred for simple finger-tapping tasks: 
We found evidence for two processes, i.e. a clock and a motor process, both of them long-range correlated.
The clock process disappeared when a metronome was used,  and the remaining motor process showed a single power law with slope around unity.
While the LRCs in both processes have similar characteristics, their origin may be very different:
When a piece of music is performed, it has structure on all scales, from motifs, phrases and themes to verses and movements. This structure is reflected in the tempo and is likely to underlie the observed long-range correlations.
Such structure is absent in finger-tapping tasks.
Those tasks are somewhat dull for the subjects, and hence it may well be that their mind is wandering during tapping. As a consequence, concentration may wax and wane, certainly on many different scales as well, thereby generating LRCs.
Studies relating neural activity to motor precision and perception hint in that direction \cite{Palva2013, Smit2013}. 
Smit et al, for example, showed clear correlation between $\beta_C$ and the scaling exponent derived from neural alpha oscillations \cite{Smit2013}. 
Hence, although the signatures of the beat time series are similar for music and finger tapping, their origin may differ vastly: One lying in the multi-scale structure of a musical composition, which aims at keeping us captivated, the other making our mind wander owing to the dullness of tapping a simple beat for minutes in a row.
Such waxing and waning of concentration and performance can be tested in future studies by simultaneously measuring markers of attention in brain activity or pupil diameter and relating this to the microtiming deviations in finger-tapping tasks; in music songs, the microtiming deviations might be related to the structure of each song.

For beat generation in music as well as in simple tapping tasks, the generative models still remain unknown. In past studies, LRCs in finger tapping were attributed either generically to models for 1/f noise, e.g. long-range correlated (critical) brain dynamics or the superposition of processes on different time scales \cite{Bak1987, Priesemann2013, Smit2013, LinkenkaerHansen2001, Beggs2003, Priesemann2014, Marom2009, Levina2017}.
Alternatively, they were explained by more mechanistic models, such as the linear phase correction model \cite{Vorberg1996}, the  shifting strategy model \cite{Wagenmakers2004}, or the hopping model \cite{West2003}, as summarized by Torre et al. \cite{Torre2009}.

We found both, the clock and the motor process to show LRCs characterized by $\beta_C \approx 0.6$ and $\beta_M \approx -1$, respectively.
Are these results for beat in music consistent with those found for finger tapping?
Early studies on finger tapping assumed that both the motor and the clock process showed uncorrelated Gaussian noise ($\beta_C \approx 0$, $\beta_M \approx -2$), but never tested that explicitly by evaluating e.g. the PSD \cite{Wing1973b,Delignieres2004}. 
First analyses of the PSD showed $0.9 < \beta_C <1.2$ for the clock process \cite{Gilden1995, Musha1985}.
The exponent $\beta_M$ was not fitted but assumed to reflect an uncorrelated process ($\beta_M \approx -2$), although the spectra were clearly flatter,  hinting at LRCs in the motor process as well. 
For the clock process, $\beta_C \approx 1$ was found consistently in various simple finger-tapping tasks \cite{Chen2002, Delignieres2004, Musha1985, Gilden1995, Smit2013}.
When two subjects tapped in synchrony, $\beta_C$ was a bit smaller ($\beta_C \approx 0.85$), and in an exemplary pop song, $\beta_C$ was found to be even smaller ($\beta_C \approx 0.56$), which is very similar to our results on the over 100 music recordings.
Overall, our study, together with the past ones, suggests that finger-tapping tasks have a larger $\beta_C$ than beat generation in music.
This indicates stronger persistence of the tempo drifts in tapping compared to music pieces.
The origin for this difference, though, remains unknown.
It is conceivable that professional musicians are better trained at keeping a constant tempo, whereas the subjects in the tapping tasks typically did not have any training. 

Regarding the motor process, results are very scarce for unpaced finger tapping.
For paced finger tapping, the $\beta_M$ were typically estimated in a different manner.
Instead of the IBI time series, the deviations from the metronome, i.e.  the error time series was used. For those, the $\beta_M'$ is expected to differ by 2, $\beta_M' = \beta_M - 2$. 
We found  $\beta_M \approx -1$, both for paced and unpaced music recordings.
For tapping, earlier studies reported $\beta_M \approx -1.5$ or  $\beta_M \approx -1.3$ \cite{Torre2008, Chen1997, Ding2002,Hennig2011}.
Hence, fluctuations  on short time scales are stronger  for tapping than for music beat generation. 
Whether these differences are attributed to the different cognitive involvements, or whether they reflect differences between lay people's tapping performance, versus professional musician's beat generation, remains an open question.

%
%
%
%
%
%

\section*{Materials and methods}
\subsection*{Datasets}
All datasets are available in the supplementary material S1 Dataset.

\subsubsection*{Dataset 1: Real-world musical performances}
\label{sec:Dataset1}

We analyzed in total 100 recordings ($47$ jazz and $53$ rock/pop), listed in Tables S2 Table -- S4 Table. 
Of these recordings, $9$ jazz and $13$ rock/pop recordings were played with metronome (``paced recordings''). 
The recordings are denoted by J** and R** with ** denoting a consecutive (arbitrary) number of the jazz or rock/pop song, respectively.  
All recordings satisfied the following criteria:
\begin{enumerate}
	\item The cymbals were clearly audible even when other high-pitched sounds were interfering.
	\item The cymbals' main rhythmic function was for pace-keeping, i.e. we discarded recordings where the cymbals only occurred occasionally or were used in a mainly expressive way, the cymbal patterns frequently changed or where drum-play was virtuoso in general. 
	\item The audio quality for MP3-encoded recordings was at least 320~kBit/s.
\end{enumerate}

\subsubsection*{Dataset 2: Experimental performances}
To obtain a complementary dataset, we asked a professional drummer to play jazz and rock/pop music as genuinely as possible on his own and in absence of a metronome.
The drummer gave informed, verbal consent that we use the recording for timing analysis.

The drummer was free in all musical decisions like tempo, rhythms and musical structure but was asked to avoid the crash cymbal and not to interrupt his performance.
Additionally he was aware of the fact that we focused on the cymbals in our analysis and thus paid attention to use them consistently.

The drummer was a professional musician with a conservatory degree in drumming, and long standing experience with live and studio jazz performances.
We obtained $7$ drum performances in total (3 jazz and 4 rock/pop) with a length of 20--30 minutes each.

A setup with six drum microphones (Shure PGA Drumkit 6) was used to record the ride cymbal, hi-hat, the toms, the snare and the bass drum separately.
In order to reduce cross-talk, we aligned each of the two overhead microphones (Shure PGA 81) to point towards the hi-hat and ride cymbal surface within a close distance ($\approx 5$~cm) and away from the other drums.

For all these recordings and performances, we estimated the beat time series as described below. 

\subsection*{Time series extraction}

In the following we detail our semi-automated workflow for millisecond-precise, reproducible beat extraction. It consists of six transformation and refinement steps in order to obtain high-precision cymbal beat time series from the initial audio signal. \\

(1) From the stereo audio signal, recorded at a sampling frequency of $44.1~$kHz, the percussion-dominant channel was isolated (Fig~\ref{fig:Pipeline}(1)).
It is denoted by $\xi(t')$, where $t'$ is the discrete time sampled at about $0.02$~ms.

(2) To detect the onsets of the cymbal, we used a time-frequency representation of $\xi(t')$. 
More specifically, we calculated the short-term Fourier transform (STFT) of  $\xi(t')$, using in the time domain a window size of 128 samples ($\approx 3$ ms), a step size of 8 samples ($\approx0.2$~ms), smoothed with a Hann function.

In the frequency domain this results in 64 bands of $f_\mathrm{Nyquist}/64 \approx 345~$Hz.
Hence for each time step $t$ (corresponding to 8 samples of $t'$) and frequency window $k$, we obtained the spectrogram $S(k,t)$. 
The cymbal was most prominent in the band from $\approx 15~$kHz to $\approx 19~$kHz. 
For higher frequencies, MP3 compression artifacts distorted the signal, and for lower frequencies, other instruments interfered.
The precise values of the frequency band depended on the specific piece and were adjusted if necessary.

(3) For every frequency band $k$, a rise in power, potentially indicating a cymbal onset, was determined by subtracting the average power of the past $9.3~$ms (i.e. 51 time steps) from the current sample:  
\begin{align*}
S'(k,t) = S(k,t) - \frac{1}{51} \sum_{i=1}^{51} S(k,t-1)
\end{align*}
The onset detection function $y(t)$ is the average of $S'(k,t)$ over the cymbal-dominated frequency bands $k$ from $\approx 15~$kHz to $\approx 19~$kHz.

(4) To extract putative cymbal events $t^\mathrm{ev}_i$, we applied a simple peak-picking algorithm by first applying a threshold ${ y_\mathrm{thresh} = 0.07\cdot\mathrm{max}\{y(t)\} }$ and then discarded all but the maximal within any time window for size $T_\mathrm{block} = 70$~ms.
This resulted in a minimum interval of $2\cdot T_\mathrm{block}$ between local maxima. 
Occasionally the threshold had to be manually lowered.

(5) To exclude all cymbal events that are not part of the beat, we first estimated the beat period $T$.
To this end, we calculated the intervals $\delta t$ between each putative cymbal event $t^\mathrm{ev}_i$ and the $m=2$ following cymbal events.
The beat period $T$ then manifested as a strong peak in a histogram of the $\delta t$ within a range [0~ms~;~1000~ms].

The local rhythmic structure of the song was obtained by plotting the $\delta t$ versus the corresponding $t^\mathrm{ev}_i$.
In this representation, the regions in the song with, e.g., fainter or missing cymbals resulted in sparsely populated regions.
For such pieces, the procedure described above was repeated with a lower threshold.
If this led to many false-detections, then we increased the default value of $m$ to $3$ or $4$.

Having estimated the beat period $T$, we  grouped the cymbal events to labeled sequences that were locally in agreement with $T$:
Two cymbal events $t^\mathrm{ev}_i$ and $t^\mathrm{ev}_j$ were assigned the same label if their time difference was within $T \pm \tau$.
$\tau$ was set to $35~$ms and adjusted if necessary.
These labeled sequences were manually assembled to full beat time series $\hat{t}_i$.
Only sequences of length 256 or longer were used for further analysis, apart from 3 slightly shorter time series (see below).

The steps described up to this point needed about one minute quality checking per recording.
They where optimized to quickly validate whether a sufficiently long sequence of beats could be extracted reliably.
In the following, we describe how for these $107$ recordings the millisecond precise onsets were extracted.

(6) First, all $\hat{t}_i$ of the putative beat time series were checked for validity and corrected manually if necessary.
Then we determined millisecond-precisely the physical onset time $t_i$ as the time were the onset detection function $y(t)$ (see step 3) first rose above base line (the blue line in panel (6) indicates the estimated physical onset time $t_i$).
$t_i$ is expected to be at most $50~$ms before the corresponding $\hat{t}_i$.
Hence starting at $\hat{t}_i-50$~ms, we scanned that entire window to find the \textit{last} $t$ for which $y(t)$ exceeded its own preceding 5~ms baseline, i.e. $y(t) > \mathrm{max}\{y(t-5~\mathrm{ms}),\dots, y(t-1~\mathrm{sample})\}$ is fulfilled.
In a few pieces, e.g. with prominent rim-shots, which result in multiple closely spaced local maxima, the most reliable type of maximum was used by defining a target interval in which $y(t_i)$ was expected to lie. Typically, the correct onset times were unambiguously visible in the spectrograms.
These automatically detected onset times $t_i$ were all checked audio-visually and adjusted if necessary.

\subsection*{Time series analysis}

We calculated the power spectral density (PSD) of the inter-beat interval (IBI) time series, i.e. the temporal difference between two successive beat onsets $d_i = t_{i+1}-t_i$.
Here, any missing $t_i$ was handled as NaN.
The IBIs were detrended with a polynomial of degree 3.
Afterwards, the NaNs were discarded, because this procedure leads to a better estimate of the PSD.
Time series with less than $256$ data points were centered and zero-padded --- this applied to three out of the 107 time series (R16, R32, R48), where with $N=194$ R48 was the shortest.

To estimate the exponents, we applied the standard Welch PSD method introduced in \cite{Welch1967} with window size $N_\mathrm{win} = 256$.
To suppress spectral leakage, each segment was multiplied with a Hann window $w(n) =\sin^2\left(\frac{\pi n}{N_\mathrm{win}-1}\right)$, where $n$ denotes the index $n=0,1,\dots, N_\mathrm{win}-1$.
The overlap was set to $N_\mathrm{overlap}\approx N_\mathrm{win}/2$, i.e. first the number of windows fitting in the time series of length $N$ was calculated and the overlap was adjusted to cover the whole time series instead of the next-smaller multiple of $N_\mathrm{win}/2$.

For unpaced performances, we fitted a V-shaped PSD (see Fig~\ref{fig:V-shape-Scheme}) using a superposition of two power laws with opposite-signed scaling parameters $\beta_C$ and $\beta_M$:
\begin{align}
\widetilde{P}(f) = P_C f^{-\beta_C} + P_M f^{-\beta_M}, \label{eq:V-Shape-Model}
\end{align}
where $\beta_C$ putatively quantifies how the clock process is correlated over time, while $\beta_M$ quantifies how the ``microtiming deviation'' from the beat or ``motor process'' is (anti)correlated over time.
$P_C$ and $P_M$ describe the power of the clock and motor components, respectively.
As the clock and the motor components have opposite sign, each of them dominates one side of the spectrum, and the turnover frequency, i.e. the resulting minimum of $P(f)$, is denoted by $f_V$ (Fig~\ref{fig:V-shape-Scheme}A).
$f_V$ is a function of the exponents and power of the motor and clock components:
\begin{align*} 
\log_2 f_V = \frac{\log_2 P_C-\log_2 P_M}{\beta_C-\beta_M}.
\end{align*}
When fitting all the free parameters $\theta = (f_V, \beta_C, \beta_M)$ of the spectrum, we aimed at weighting the motor and clock contributions equally.
To this end, we assumed a turn-over frequency $f_V^*$ and weighted the fit residuals of both sides of $f_V^*$  equally.
This results in a weighting function

\begin{align*}
w(f; f_V^*) = 
\begin{cases}
1/N^-, & \text{if~} f \leq f_V^* \\
1/N^+, & \text{if~} f > f_V^*
\end{cases}
\end{align*}
where $N^-$ ($N^+$) denotes the number of frequency bins $f$ that are smaller (larger) than $f_V^*$.
The residual
\begin{align*}
r(\theta) = \sum\limits_{f}  w(f; f_V^*) |\log_2P(f)-\log_2\widetilde{P}(f; \theta)|
\end{align*}
between model and data was minimized using the Broyden-Fletcher-Goldfarb-Shanno algorithm \cite{Broyden1970}.
To initialize the minimization, for each trial each side of the log-transformed spectrum was approximated by a linear relationship, using the Theil-Sen method \cite{Birkes1993}. 
As $f_V^*$ is not known a priori, we scanned  $\log_2 f_V^*$ equidistantly on ${\log_2 f_V^*\in[-6;-2]}$.
For each time series we thus obtained ten parameter sets $\theta$.
Note that $f_V^*$ was only used to define the weighting function $w(f)$, while $f_V$ proper is a free parameter.
We still used different $f_V^*$, because it allowed for an estimate of the variability of the parameters $\theta$.

In the case of the paced (metronome-guided) recordings we expected the clock component to be missing.
As a consequence, the PSD approximates a power law with $P(f) \sim f^{-\beta_M}$.
Thus when fitting with the procedure above, the parameters $\beta_M$ and $\beta_C$ are expected to be both negative.
This condition was used as a test for paced versus unpaced pieces.
The $\beta_M$ of the unpaced pieces can then be estimated either by fitting a single power law, or by fitting the V-shape as above.

\subsection*{Permutation test and effect size}
\label{sec:Permutation_Test}
To test for the presence of different effects for jazz (J) versus rock/pop (R), we applied a median-based two-sided permutation test on the estimated parameters $\beta_C$, $\beta_M$ and $\log_2 f_V$ obtained by the V-shaped fit.
Therefore, the median values ${\bar{x} = \{\bar{\beta}_C, \bar{\beta}_M, \log_2\bar{f_V}\}}$ for jazz and rock/pop were compared by $\Delta x = \bar{x}_\mathrm{J}-\bar{x}_\mathrm{R}$ for each parameter.

In addition to the p-values from the permutation, we reported the effect size for the differences between jazz and rock/pop recordings.
We used a modified Cohen's $d$
\begin{align*}
d = \frac{\bar{x}_\mathrm{J}-\bar{x}_\mathrm{R}}{\sigma},
\end{align*}
where $\bar{x}$ denotes the median of the respective values $\bar{\beta}_C$, $\bar{\beta}_M$ or $\log_2 \bar{f_V}$.
The pooled standard deviation $\sigma$ was computed from the population sizes $n_\mathrm{J}$, $n_\mathrm{R}$ and standard deviations $\sigma_\mathrm{J}$, $\sigma_\mathrm{R}$ of the respective populations:
\begin{align*}
\sigma = \sqrt{\frac{(n_\mathrm{J}-1)\sigma_\mathrm{J}^2+(n_\mathrm{R}-1)\sigma_\mathrm{R}^2}{n_\mathrm{J}+n_\mathrm{R}-2}}.
\end{align*}

Effect sizes are considered being small  for $0.2<|d| <0.5$, medium for $0.5 < |d| < 0.8$ and large  for $|d| > 0.8$.


\section*{Supporting information}

\textbf{S1 Dataset. All time series of beat onsets.}\\
\textbf{S2 Table. List of unpaced jazz performances.}\\
\textbf{S3 Table. List of unpaced rock/pop performances.}\\
\textbf{S4 Table. List of paced jazz and rock/pop performances.}

%

\section*{Acknowledgments}
We thank Annette Witt and George Datseris for helpful discussions and support, and Tim Dudek for making studio drum recordings for us. 

%
%
%


\clearpage

\end{document}